\documentclass[12pt]{article}

\usepackage{times}

\usepackage{amsmath}
\usepackage{amssymb}
\usepackage{amsfonts}
\usepackage{latexsym}
\usepackage{theorem}
\usepackage{fullpage}

\usepackage{url}
\usepackage[breaklinks,bookmarks]{hyperref}

\newtheorem{theorem}{Theorem}
\newtheorem{lemma}[theorem]{Lemma}
\newtheorem{fact}[theorem]{Fact}
\newtheorem{corollary}[theorem]{Corollary}
\newtheorem{definition}[theorem]{Definition}

\def\squareforqed{\hbox{\rlap{$\sqcap$}$\sqcup$}}
\def\qed{\ifmmode\squareforqed\else{\unskip\nobreak\hfil
\penalty50\hskip1em\null\nobreak\hfil\squareforqed
\parfillskip=0pt\finalhyphendemerits=0\endgraf}\fi}
\newenvironment{proof}{\begin{trivlist}\item[]{\flushleft\bf Proof }}
{\qed\end{trivlist}}

\newtheorem{example}{Example: Ordered Seaching}

\newcommand{\ket}[1]{| #1 \rangle}
\newcommand{\bra}[1]{\langle #1 |}
\newcommand{\braket}[2]{\langle #1 | #2 \rangle}
\newcommand{\braopket}[3]{\langle #1 | #2 | #3 \rangle}

\newcommand{\zo}{\{0,1\}}
\newcommand{\adv}{\textup{Adv}}
\newcommand{\cert}{\textup{C}}
\newcommand{\bs}{\textup{bs}}
\newcommand{\thr}{\textsf{Thr}}
\newcommand{\searchGamma}{\Gamma^{\textup{search}}}

\newcommand{\nth}[1]{{\ensuremath{{#1}^{\text{th}}}}}
\newcommand{\nst}[1]{{\ensuremath{{#1}^{\text{st}}}}}
\newcommand{\op}[1]{\mathsf{#1}}

\newcommand{\myatop}[2]{\genfrac{}{}{0pt}{}{#1}{#2}}

\newcommand{\reals}{\Re}

\renewcommand{\deg}{\mathrm{deg}}
\newcommand{\adeg}{%
\widetilde{\vphantom{\raisebox{-.05cm}{d}}\smash{\mathrm{deg}}}}
\newcommand{\rdeg}{\mathrm{rdeg}}

\title{Lower Bounds on Quantum Query Complexity}
\author{%
  Peter H{\o}yer%
  \thanks{Department of Computer Science,
  University of Calgary.
  Supported by Canada's Natural Sciences and Engineering 
  Research Council (NSERC),
  the Canadian Institute for Advanced Research (CIAR), 
  and The Mathematics of Information Technology and Complex 
  Systems (MITACS).}\\
  {\tt hoyer@cpsc.ucalgary.ca}
\and
Robert {\v S}palek\thanks{%
  CWI and University of Amsterdam.  
  Supported in part by the EU fifth framework project RESQ, 
  IST-2001-37559.  
  Work conducted in part while visiting the University of Calgary.}\\
  {\tt sr@cwi.nl}
}
\date{September 2005}

\begin{document}
\maketitle

\begin{abstract}
Shor's and Grover's famous quantum algorithms for factoring and
searching show that quantum computers can solve certain computational
problems significantly faster than any classical computer.  We discuss
here what quantum computers \emph{cannot} do, and specifically how to
prove limits on their computational power.  We cover the main known
techniques for proving lower bounds, and exemplify and compare the
methods.  
\end{abstract}

\section{Introduction}
\label{sec:intro}
The very first issue of the Journal of the ACM was published in
January 1954.  It was the first journal devoted to computer science.
For its 50th anniversary volume, published in January 2003,
editors-in-chief Joseph Y.~Halpern asked winners of the Turing Award
and the Nevanlinna Prize to discuss up to three problems that they
thought would be major problems for computer science in the next 50
years.  Nevanlinna Prize winner Leslie G.~Valiant~\cite{valiant:2003}
describes three problems, the first of which is on physically
realizable models for computation and formalizes the setting by
defining: ``We therefore call our class PhP, the class of physically
constructible polynomial resource computers.''  He then formulates the
problem by: ``[t]o phrase a single question, the full characterization
of PhP,'' and argues that ``this single question appears at this time
to be scientifically the most fundamental in computer science.''

On~January~26, this year, Nobel Laureate David Gross gave a CERN
Colloquium presentation on ``The future of
physics''~\cite{gross:2005}.  He discusses ``25 questions that might
guide physics, in the broadest sense, over the next 25 years,'' and
includes as questions 15 and~16 ``Complexity'' and ``Quantum
Computing.''  In~July, this year, the Science magazine celebrated its
125th anniversary by ``explor[ing] 125 big questions that face
scientific enquiry over the next
quarter-century''~\cite{science:2005}.  Among the top~25, is the
question of ``What are the limits of conventional computing?''
Charles Seife writes: ``[T]here is a realm beyond the classical
computer: the quantum,'' and he discusses the issue of determining
``what quantum-mechanical properties make quantum computers so
powerful.''

In this issue of the Bulletin of the EATCS, we would like to offer an
introduction to the topic of studying limitations on the power of
quantum computers.  Can quantum computers really be more powerful than
traditional computers?  What can quantum computers not do?  What proof
techniques are used for proving bounds on the computational power of
quantum computers?  It is a highly active area of research and
flourishing with profound and beautiful theorems.  Though deep, it is
fortunately also an accessible area, based on basic principles and
simple concepts, and one that does not require specialized prior
knowledge.  One aim of this paper is to show this by providing a
fairly complete introduction to the two most successful methods for
proving lower bounds on quantum computations, the adversary method and
the polynomial method.  Our survey is biased towards the adversary
method since it is likely the least familiar method and it yields very
strong lower bounds.  This paper is meant to be supplemented by the
excellent survey of Buhrman and
de~Wolf~\cite{buhrman&wolf:dectreesurvey} on decision tree
complexities, published in 2002 in the journal Theoretical Computer
Science.

We demonstrate the methods on a running example, and for this, we use
one of the most basic algorithmic questions one may think of: that of
searching an ordered set.  Can one implement ordered searching
significantly faster on a quantum computer than applying a standard
$\Theta(\log N)$ binary search algorithm?

The rest of the paper is organized as follows.  We motivate and define
our models of computation in the next section.  We then discuss very
basic principles used in proving quantum lower bounds in
Section~\ref{sec:distinguish} and use them to establish our first
lower-bound method, the adversary method, in
Section~\ref{sec:adversary}.  We discuss how to apply the method in
Section~\ref{sec:applying}, and its limitations in
Section~\ref{sec:limits}.  We then give an introduction to the second
method, the polynomial method, in Section~\ref{sec:poly}.  We compare
the two methods in Section~\ref{sec:applyingpoly} and give a few final
remarks in Section~\ref{sec:concluding}.

We have aimed at limiting prior knowledge on quantum computing to a
bare minimum.  Sentences and paragraphs with kets and bras
($\ket{{\text{this is a ket}}}$ and $\bra{{\text{this is a bra}}}$)
can either safely be skipped, or substituted with column-vectors and
row-vectors, respectively.

\section{Quantum query complexity}
\label{sec:query}
Many quantum algorithms are developed for the so-called oracle model
in which the input is given as an oracle so that the only knowledge we
can gain about the input is in asking queries to the oracle.  The
input is a finite bitstring $x \in \zo^N$ of some length $N$, where $x
= x_1 x_2 \ldots x_N$.  The goal is to compute some function $F:\zo^N
\rightarrow \zo^m$ of the input~$x$.  Some of the functions we
consider are boolean, some not.  We use the shorthand notation $[N] =
\{1,2,\ldots, N\}$.

As our measure of complexity, we use the query complexity.  The query
complexity of an algorithm $\op{A}$ computing a function~$F$ is the
number of queries used by~$\op{A}$.  The query complexity of $F$ is
the minimum query complexity of any algorithm computing~$F$.  We are
interested in proving lower bounds on the query complexity of specific
functions and consider methods for computing such lower bounds.

An alternative measure of complexity would be to use the time
complexity which counts the number of basic operations used by an
algorithm.  The time complexity is always at least as large as the
query complexity since each query takes one unit step, and thus a
lower bound on the query complexity is also a lower bound on the time
complexity.  For most existing quantum algorithms, including Grover's
algorithm~\cite{grover:search}, the time complexity is within
poly-logarithmic factors of the query complexity.  A~notorious
exception is the so-called Hidden Subgroup Problem which has
polynomial query complexity~\cite{ehk:subgroup}, yet polynomial time
algorithms are known only for some instances of the problem.

The oracle model is called decision trees in the classical setting.
A~classical query consists of an index $i \in [N]$, and the answer of
the bit~$x_i$.  There is a natural way of modelling a query so that it
is reversible.  The input is a pair $(i,b)$, where $i\in [N]$ is an
index and $b \in \zo$ a bit.  The output is the pair $(i,b \oplus
x_i)$, where the bit $b$ is flipped if $x_i = 1$.  There are (at
least) two natural ways of generalizing a query to the quantum
setting, in which we require all operations to be unitary.  The first
way is to consider a quantum query as a unitary operator that takes
two inputs $\ket{i}\ket{b}$, where $i \in [N]$ and $b \in \zo$, and
outputs $\ket{i}\ket{b \oplus x_i}$.  The oracle is then simply just a
linear extension of the reversible query given above.  We extend the
definition of the oracle so that we can simulate a non-query, and we
allow it to take some arbitrary ancilla state $\ket{z}$ with $z \geq
0$ as part of the input and that is acted upon trivially,
\begin{equation}
\op{O}'_x \ket{i,b;z} = \begin{cases}
\ket{i,b ;z} & \text{ if $i=0$ or $x_i =0$}\\
\ket{i,b \oplus 1;z} & \text{ if $i \in [N]$ and $x_i =1$.}
	      \end{cases}
\end{equation}
The ancilla $\ket{z}$ contains any additional information currently
part of the quantum state that is not involved in the query.

The second way is to consider a quantum query as a unitary operator
${\op{O}}_x$ that takes only the one input $\ket{i}$ and outputs
$(-1)^{x_i} \ket{i}$, where $i \in [N]$.  We say that the oracle is
``computed in the phases'' by~${\op{O}}_x$.  Both operators
$\op{O}'_x$ and ${\op{O}}_x$ square to the identity, i.e., they are
their own inverses, and thus unitary.  The two operators are equivalent
in that one query to either oracle can be simulated by a superposition
query to the other oracle preceeded and followed by a basis change.
Though the first
way is possibly the more intuitive, we shall adapt the second way as
it is very convenient when proving lower bounds.  Again, we extend the
definition of the oracle ${\op{O}}_x$ so that it also embodies a
non-query, and we allow it to take some arbitrary ancilla state
$\ket{z}$ that is not acted upon,
\begin{equation}
{\op{O}}_x \ket{i;z} = \begin{cases}
\ket{i;z} & \text{ if $i=0$}\\
(-1)^{x_i} \ket{i;z} & \text{ if $1 \leq i \leq N$.}
	      \end{cases}
\end{equation}

We may think of one query as a one-round exchange of information
between two parties, the algorithm and the oracle.  In the classical
setting, the algorithm sends an index $i \in [N]$ to the oracle, and
the oracle responds with one bit of information, namely~$x_i$.  In the
quantum setting, the algorithm sends the $\log_2(N)$ qubits $\ket{i}$
to the oracle ${\op{O}}_x$, and the oracle responds with $(-1)^{x_i}
\ket {i}$.  The algorithm and oracle thus exchange a total number of
$2 \log_2(N)$ qubits, and thus, a quantum query to ${\op{O}}_x$ can
convey up to $2 \log_2(N)$ classical bits of information about the
oracle by Holevo's theorem~\cite{holevo:comm,cdnt:ip} and
superdense coding~\cite{bw:superdense}.

Information theoretically, a function $F: \zo^N \rightarrow
\zo^{\log_2(N)}$ that outputs at most $O(\log_2(N))$ bits, can
potentially be solved by a constant number of queries to the oracle.
An example of such a problem is the Deutsch-Jozsa
problem~\cite{dj:rapid}, which is to distinguish balanced boolean
functions from constant functions.  (A~function $F$ is constant if
$F(x)=F(y)$ for all inputs $x, y$, and it is balanced if it is not
constant and $|F^{-1}(F(x))| =|F^{-1}(F(y))|$ for all inputs $x, y$.)

A~quantum algorithm in the oracle model starts in a state that is
independent of the oracle.  For convenience, we choose the state
$\ket{0}$ in which all qubits are initialized to~$0$.  It then evolves
by applying arbitrary unitary operators $\op{U}$ to the system,
alternated with queries $\op{O}_x$ to the oracle $x$, followed by a
conclusive measurement of the final state, the outcome of which is the
result of the computation.  In~symbols, a~quantum algorithm $\op{A}$
that uses $T$ queries, computes the final state
\begin{equation}
\ket{\psi^T_x} = {\op{U}}_T {\op{O}}_x {\op{U}}_{T-1} \cdots
{\op{U}}_1 {\op{O}}_x {\op{U}}_0 \ket{0}
\end{equation}
which is then measured.  If the algorithm computes some function $F:
\zo^N \rightarrow \zo^m$, we measure the $m$ leftmost bit of the final
state $\ket{\psi^T_x}$, producing some outcome~$w$.  The success
probability $p_x$ of $\op{A}$ on input $x \in \zo^N$ is the
probability that $w = F(x)$.  For complete functions $F:\zo^N
\rightarrow \zo^m$, we define the success probability of~$\op{A}$ as
the minimum of $p_x$ over all $x \in \zo^N$.  For partial functions
$F:S \rightarrow \zo^m$, where $S \subseteq \zo^N$, we take the
minimum over~$S$ only.  A~quantum algorithm $\op{A}$ has error at most
$\epsilon$ if the success probability of $\op{A}$ is at
least~$1-\epsilon$.  Let $Q_\epsilon(F)$ denote the minimum query
complexity of any quantum algorithm that computes $F$ with two-sided
error at most~$\epsilon$, and as common, let $Q_2(F) = Q_{1/3}(F)$
denote the two-sided bounded error complexity with $\epsilon=1/3$.

As our running example, we use the well-known ordered searching
problem.  In the oracle model, the input to ordered searching is an
$N$-bit string $x = (x_1, \ldots, x_{N})$.  We are promised that $x_i
\leq x_{i+1}$ for all $1 \leq i <N$ and that $x_N=1$, and the goal is
to find the leftmost~1, i.e., the index $i \in [N]$ for which $x_i =1$
and no index $j<i$ exists with $x_j=1$.
\begin{description}
\item[\quad Given:] An $N$-bit string $x = (x_1, x_2, \ldots, x_{N})$
given as an oracle.
\item[\quad Promise:] $x_{i} \leq x_{i+1}$ for $1 \leq i < N$
and $x_N = 1$.
\item[\quad Output:] Index $i$ such that $x_i=1$ and either $x_{i-1}=0$ or
  $i=1$.
\end{description}
The classical query complexity of ordered searching is $\lceil
\log_2(N) \rceil$ and is achieved by standard binary searching.  The
quantum query complexity is at most $0.45 \log_2 N$, due to the work
of high school student M.~B.{} Jacokes in collaboration with Landahl
and Brookes~\cite{jlb:search} (See also
\cite{fggs99:search,hns:ordered-search}).  Using the adversary method,
we show that their algorithm is within a factor of about two of being
optimal.

\section{Distinguishing hard inputs}
\label{sec:distinguish}
The first quantum lower bound using adversary arguments was given by
Bennett, Bernstein, Brassard, and Vazirani in~\cite{bbbv:hybrid}.
They show that any quantum query algorithm can be sensitive to at most
quadratically many oracle bits, which implies a lower bound of
$\Omega(\sqrt{N})$ for Grover's problem~\cite{grover:search} and thus
proves that Grover's $O(\sqrt{N})$ algorithm is optimal.  Grover's
problem is a search problem in which we are given an $N$-bit string $x
\in \zo^N$ as an oracle, and the goal is to find an index $i$ for
which $x_i=1$, provided one exists.  Interestingly, the lower bound of
Bennett et~al.{} was proved in 1994, well before Grover defined his
search problem.  In~2000, Ambainis~\cite{ambainis:lowerb} found an
important generalization of the method and coined it ``adversary
arguments.''

A~constructive interpretation of basic adversary arguments is in terms
of \emph{distinguishability}.  We will thus not be concerned with
computing the function~$F$, but merely interested in {distinguishing}
oracles.  Consider some algorithm $\op{A}$ that computes some
function~$F$ in the oracle model, and consider two inputs $x, y \in
\zo^N$ for which $F(x) \neq F(y)$.  Since $\op{A}$ computes $F$, it
must in particular be capable of distinguishing between oracle $x$ and
oracle~$y$.  For a given problem we try to identify \emph{pairs of
oracles} that are hard to \emph{distinguish}.  If we can identify hard
input pairs, we may derive a good lower bound.  However, a caveat is
that using only the very hardest input pairs does not yield good lower
bounds for some problems, and we are thus naturally led to also
consider less hard input pairs.  A~remedy is to use \emph{weights}
that capture the hardness of distinguishing each pair of oracles, and
to do so, we define a matrix $\Gamma$ of dimension $2^N \times 2^N$
that takes non-negative real values,
\begin{equation}
\Gamma : \zo^N \times \zo^N \rightarrow \reals^{+}_0.
\end{equation}
We require that $\Gamma$ is symmetric and that $\Gamma[x,y] = 0$
whenever $F(x)=F(y)$.  We say that $\Gamma$ is a \emph{spectral
adversary matrix for $F$} if it satisfies these two conditions.  The
symmetry condition on $\Gamma$ states that we are concerned with
distinguishing \emph{between} any two inputs $x,y$.  We are not
concerned with distinguishing $x$ \emph{from} $y$, nor distinguishing
$y$ \emph{from}~$x$.  We discuss this subtlety further in
Section~\ref{sec:applying} below when considering alternative
definitions of weighted adversary arguments.  The spectral adversary
matrix $\Gamma$ allows us to capture both total and partial functions,
as well as non-boolean functions.  Since we are only concerned with
distinguishability, once we have specified the entries of $\Gamma$, we
may safely ignore the underlying function~$F$.

Weighted adversary arguments were first used by H{\o}yer, Neerbek, and
Shi in~\cite{hns:ordered-search} to prove a lower bound of
$\Omega(\log N)$ for ordered searching and $\Omega(N \log N)$ for
sorting.  Barnum and Saks~\cite{bs:q-read-once} used weighted
adversary arguments to prove a lower bound of $\Omega(\sqrt{N})$ for
read-once formulae, and introduced the notion~$\Gamma$ that we adapt
here.  Barnum, Saks, and Szegedy extended their work
in~\cite{bss:semidef} and derived a general lower bound on the query
complexity of~$F$ in terms of spectral properties of matrix~$\Gamma$.
Their lower bound has a very elegant and short formulation, a basic
proof, and captures important properties of adversary methods, and we
shall thus adapt much of their terminology.

As discussed above, the key to prove a good lower bound is to pick a
good adversary matrix~$\Gamma$.  For our running example of ordered
searching, which is a partial non-boolean function, we use the
following weights.
\begin{example}
The weight on the pair $(x,y)$ is the inverse of the Hamming distance
of $x$ and~$y$,
\begin{equation}
\searchGamma[x,y] =
\begin{cases}
 \frac{1}{|F(x) - F(y)|} & \textup{ if $x$ and $y$ are
  valid and distinct inputs to~$F$}\\
 0 & \textup{ otherwise.}
\end{cases}
\end{equation}
The larger the Hamming distance between $x$ and~$y$, the easier it is to
distinguish them, and the smaller weight is assigned to the pair.
\end{example}

We have to choose how to measure distinguishability.  The possibly
simplest measure is to use inner products.  Two quantum states are
distinguishable with certainty if and only if they are orthogonal, and
they can be distinguished with high probability if and only if their
inner product has small absolute value.

\begin{fact}\label{fact:disting}
Suppose we are given one of two known states $\ket{\Psi_x},
\ket{\Psi_y}$.  There exists a measurement that correctly determines
which of the two states we are given with error probability at most
$\epsilon$ if and only if $|\braket{\Psi_x}{\Psi_y}| \leq \epsilon'$,
where $\epsilon' = 2\sqrt{\epsilon(1-\epsilon)}$.
\end{fact}

Since a unitary operator is just a change of basis, it does not change
the inner product between any two quantum states, and thus the inner
product can only change as a consequence of queries to the oracle.

\section{Adversary lower bounds}
\label{sec:adversary}

Adversary lower bounds are of information theoretical nature.  A~basic
idea in adversary lower bounds is to upper-bound the amount of
information that can be learned in a single query.  If little
information can be learned in any one query, then many queries are
required.  We use spectral properties of~$\Gamma$ to put an upper
bound on the amount of information the algorithm learns about the
oracle.

Let $\op{A}$ be some quantum algorithm that computes some function $F$
with bounded two-sided error.  For every integer $t \geq 0$ and every
oracle~$x$, let
\begin{equation}
\ket{\psi_x^t} = {\op{U}}_t {\op{O}}_x \cdots {\op{U}}_1 {\op{O}}_x
{\op{U}}_0 \ket{0}
\end{equation}
denote the quantum state after $t$ queries to the oracle.  To measure
the progress of the algorithm, we define similarly to~\cite{
ambainis:lowerb, hns:ordered-search, bs:q-read-once, bss:semidef} a
weight function
\begin{equation}
W^t = \sum_{x,y} \Gamma[x,y] \delta_x \delta_y \cdot
\braket{\psi_x^t}{\psi_y^t},
\end{equation}
where $\delta$ is a fixed principal eigenvector of~$\Gamma$, i.e., a
normalized eigenvector corresponding to the largest eigenvalue
of~$\Gamma$, and where $\delta_x$ denotes the \nth{x} entry
of~$\delta$.

The algorithm starts in a quantum state $\ket{\psi_x^0} = {\op{U}}_0
\ket{0}$ which is independent of the oracle~$x$, and thus the total
initial weight is
\begin{equation}
W^0 = \sum_{x,y} \Gamma[x,y] \delta_x \delta_y = \lambda(\Gamma),
\end{equation}
where $\lambda(\Gamma)$ denotes the spectral norm of~$\Gamma$.  The
final state of the algorithm after $T$ queries is $\ket{\psi_x^T}$ if
the oracle is $x$, and it is $\ket{\psi_y^T}$ if the oracle is~$y$.
If $F(x) \neq F(y)$, we must have that $|\braket{\psi^T_x}{\psi^T_y}|
\leq \epsilon'$ by Fact~\ref{fact:disting}, and hence $W^T \leq
\epsilon' W^0$.  If the total weight can decrease by at most $\Delta$
by each query, the algorithm requires $\Omega(\frac{W^0}{\Delta})$
queries to the oracle.

Following Barnum, Saks, and Szegedy~\cite{bss:semidef}, we upper
bound~$\Delta$ by the largest spectral norm of the
matrices~$\Gamma_i$, defined by
\begin{equation}
\Gamma_i[x,y] = 
\begin{cases}
\Gamma[x,y] & \text{ if $x_i \neq y_i$}\\
            0 & \text{ if $x_i = y_i$,}
\end{cases}
\end{equation}
for each $1 \leq i \leq n$.  The theorem of~\cite{bss:semidef} is here
stated (and proved) in a slightly more general form than
in~\cite{bss:semidef} so that it also applies to non-boolean
functions.  Our proof aims at emphasizing distinguishability and
differs from the original.

\begin{theorem}[Spectral method~\cite{bss:semidef}]
\label{thm:spectral}
For any adversary matrix $\Gamma$ for any function $F: \zo^N
\rightarrow \zo^m$,
\begin{equation}
Q_2(F) = \Omega\Big(\frac{\lambda(\Gamma)} {\max_i
\lambda(\Gamma_i)}\Big).
\end{equation}
\end{theorem}

\begin{proof}
We prove that the drop in total weight $W^t - W^{t+1}$ by the
\nst{t+1} query is upper-bounded by the largest eigenvalue of the
matrices~$\Gamma_i$.

For each $0 \leq i \leq N$, let $\op{P}_i = \sum_{z \geq 0}
\ket{i;z}\bra{i;z}$ denote the projection onto the subspace querying
the \nth{i} oracle bit.  Let $\beta_{x,i} = | \op{P}_i
\ket{\psi_x^t}|$ denote the absolute value of the amplitude of
querying the \nth{i} bit in the \nst{t+1} query, provided the oracle
is~$x$.  Note that $\sum_{i=0}^N \beta_{x,i}^2 = 1$ for any
oracle~$x$, since the algorithm queries one of the $N$ bits $x_1,
\ldots, x_N$, or simulates a non-query by querying the oracle with
$i=0$.  The \nst{t+1} query changes the inner product by at most the
overlap between the projections of the two states onto the subspace
that corresponds to indices~$i$ on which $x_i$ and $y_i$ differ,
\begin{equation}\label{eq:pair}
\Big| 
\braket{\psi_x^t}{\psi_y^t} - \braket{\psi_x^{t+1}}{\psi_y^{t+1}} 
\Big|
=
\Big| 
\braopket{\psi_x^t}{(\op{I} - \op{O}_x \op{O}_y)}{\psi_y^t} 
\Big|
= \Big| 2\sum_{i: x_i \neq y_i} 
  \braopket{\psi_{x}^{t}}{\op{P}_i}{\psi_{y}^{t}} \Big|
\leq 2 \sum_{i: x_i \neq y_i} \beta_{x,i}\beta_{y,i}.
\end{equation}
The bigger the amplitudes of querying the bits $i$ on which $x_i$ and $y_i$
differ, the larger the drop in the inner product can be.

Define an auxiliary vector $a_i[x] = \delta_x \beta_{x,i}$ and note
that
\begin{equation*}
\sum_{i=0}^N |a_i|^2 = \sum_{i=0}^N \sum_x \delta_x^2 \beta_{x,i}^2 =
\sum_x \delta_x^2 \sum_{i=0}^N \beta_{x,i}^2 = \sum_x \delta_x^2 = 1.
\end{equation*}
The drop in the total weight is upper-bounded by
\begin{eqnarray*}
\big| W^t - W^{t+1} \big|
&=& \Big| \sum_{x,y} \Gamma[x,y] \delta_x \delta_y
  \big( \braket{\psi_x}{\psi_y} - \braket{\psi'_x}{\psi'_y}
  \big) \Big| \\
&=& \Big| 2 \sum_{x,y} \sum_{i: x_i\ne y_i} \Gamma[x,y] \delta_x \delta_y
  \braopket{\psi_{x}}{\op{P}_i}{\psi_{y}} \Big| \\
&\leq& 2 \sum_{x,y} \sum_i \Gamma_i[x,y] \delta_x \delta_y
  \cdot \beta_{x,i} \beta_{y,i} \\
&=& 2 \sum_i a_i^* \Gamma_i a_i \\
&\leq& 2 \sum_i \lambda(\Gamma_i) |a_i|^2 \\
&\leq& 2 \max_i \lambda(\Gamma_i) \cdot \sum_i |a_i|^2\\
&=&  2 \max_i \lambda(\Gamma_i).
\end{eqnarray*}
Here $a_i^*$ denotes the transpose of~$a_i$.  The first inequality
bounds the drop in inner product for a specific pair and follows from
Equation~\ref{eq:pair}.  The second inequality follows from the
spectral norm of~$\Gamma$.  The second and third inequalities state
that the best possible query distributes the amplitude of the query
according to the largest principal eigenvector of the query
matrices~$\Gamma_i$.
\end{proof}

\begin{example}
Returning to our example of ordered searching, for \mbox{$N=4$}, the
adversary matrix with respect to the ordered basis $(0001, 0011, 0111,
1111)$ is given~by
\begin{equation*}
{\searchGamma}^{(4)} = \left[\begin{matrix}
0 & 1 & \frac{1}{2} & \frac{1}{3} \\
1 & 0 & 1 & \frac{1}{2}\\
\frac{1}{2} & 1 & 0 & 1\\
\frac{1}{3} & \frac{1}{2} & 1 & 0
\end{matrix}\right].
\end{equation*}
The spectral norm is easily seen to be lower-bounded by the sum of the
entries in the first row, $\lambda({\searchGamma}^{(4)}) \geq 1 +
\frac{1}{2} + \frac{1}{3}$.  In general, $\lambda(\searchGamma)$ is
lower-bounded by the harmonic number $H_{N-1}$, which is at least
$\ln(N)$.  The spectral norm of the query matrices
$\lambda(\searchGamma_i)$ is maximized when $i= \lfloor N/2\rfloor$,
in which case it is upper-bounded by the spectral norm of the infinite
Hilbert matrix $[1/(r+s-1)]_{r,s \geq 1}$, which is~$\pi$.  We thus
reprove the lower bound of $(1-\epsilon') \frac{\ln(N)}{\pi}$ for
ordered searching given in~\cite{hns:ordered-search}.
\end{example}

\section{Applying the spectral method}
\label{sec:applying}

The spectral method is very appealing in that it has a simple
formulation, a basic proof, and gives good lower bounds for many
problems.  {\v S}palek and Szegedy~\cite{ss:adversary} show that for any
problem, the best lower bound achievable by the spectral method is
always at least as good as the best lower bound achievable by any of
the previously published adversary methods.  Their proof is
constructive and illuminating: given any lower bound in any of the
previously published adversary methods, they construct an adversary
matrix~$\Gamma$ and prove it achieves the same lower bound.

The first general quantum lower bound using adversary arguments was
introduced by Ambainis in~\cite{ambainis:lowerb}.  As shown
in~\cite{ss:adversary}, it can be derived from the spectral method by
applying simple bounds on the spectral norm of $\Gamma$ and
each~$\Gamma_i$.  By definition, the numerator $\lambda(\Gamma)$ is
lower-bounded by $\frac{1}{|d|^2} d^* \Gamma d$ for any non-negative
vector~$d$, and by Mathias' lemma~\cite{mathias:spectral-norm}, the
denominator $\lambda(\Gamma_i)$ is upper-bounded by the product of a
row-norm and a column-norm.

\begin{lemma}[\cite{mathias:spectral-norm, ss:adversary}]
\label{lm:mathias}
Let $G$ be any non-negative symmetric matrix and $M, N$ non-negative
matrices such that $G = M \circ N$ is the entrywise product of $M$
and~$N$.  Then
\begin{equation*}
\lambda(G) \leq \max_{\myatop{x,y} {G[x,y] > 0}} r_x(M) \; c_y(N),
\end{equation*}
where $r_x(M)$ is the $\ell_2$-norm of the \nth x row in $M$, and
$c_y(N)$ is the $\ell_2$-norm of the \nth y column in~$N$.
\end{lemma}

Applying these two bounds, we obtain Ambainis' lower bound
in~\cite{ambainis:lowerb}.  We refer to the method as an unweighted
adversary method since it considers only two types of inputs: easy
inputs and hard inputs.  We construct a zero-one valued adversary
matrix $\Gamma$ that corresponds to a uniform distribution over the
hard input pairs.

\begin{theorem}[Unweighted method \cite{ambainis:lowerb}]
\label{thm:unweighted}
Let $F$ be a partial boolean function, and let $A \subseteq F^{-1}(0)$
and $B \subseteq F^{-1}(1)$ be subsets of (hard) inputs.  Let $R
\subseteq A \times B$ be a relation, and set $R_i = \{ (x,y) \in R:
x_i \neq y_i \}$ for each $1 \leq i \leq n$.  Let $m, m'$ denote the
minimal number of ones in any row and any column in relation $R$,
respectively, and let $\ell, \ell'$ denote the maximal number of ones
in any row and any column in any of the relations~$R_i$, respectively.
Then $Q_2(f) = \Omega(\sqrt{m m' / \ell \ell'})$.
\end{theorem}

\begin{proof}
Let $S = \{ (x,y): (x,y) \in R \vee (y,x) \in R \}$ be a symmetrized version
of $R$.  Define a column vector $d$ from the relation $S$ by setting $d_x
= \sqrt{|\{ y: (x,y) \in S \}|}$, and an adversary matrix $\Gamma$ by setting
$\Gamma[x,y]=\frac{1}{d_x d_y}$ if and only if $(x,y) \in S$.  Then
$\lambda(\Gamma) \geq \frac{1}{|d|^2} d^* \Gamma d = 1$.  For each of the
matrices $\Gamma_i$, we apply Lemma~\ref{lm:mathias} with $M[x,y] = N[y,x] =
\frac{1}{d_x}$ if and only if $(x,y) \in S$.  
For every $(x,y) \in R$, $r_x(M) \le
\sqrt{\ell / d_x^2} \le \sqrt{\ell / m}$ and $c_y(N) \le \sqrt{\ell' /
d_{\vphantom{x}\smash{y}}^2} \le \sqrt{\ell' / m'}$.  For every $(x,y)
\in S-R$, the two
inequalities are swapped.  By Lemma~\ref{lm:mathias}, $\lambda(\Gamma_i) \leq
\max_{x,y: \Gamma_i[x,y] > 0} r_x(M) c_y(N) \le \sqrt{\ell \ell' / m m'}$.
\end{proof}

The unweighted adversary method is very simple to apply as it requires
only to specify a set $R$ of hard input pairs.  It gives tight lower
bounds for many computational problems, including inverting a
permutation~\cite{ambainis:lowerb}, computing any symmetric function
and counting~\cite{nw:median, bcwz:qerror, bhmt:countingj}, constant-level
and-or trees~\cite{ambainis:lowerb, hmw:berror-search}, and various
graph problems~\cite{dhhm:graph}.  For some computational problems,
the hardness does however not necessarily rely only on a few selected
hard instances, but rather on more global properties of the inputs.
Applying the unweighted method on ordered searching would for instance
only yield a lower bound of a constant.  In these cases, we may apply
the following weighted variant of the method, due to
Ambainis~\cite{ambainis:degree-vs-qc} and Zhang~\cite{zhang:ambainis}.

\begin{theorem}
[Weighted method \cite{ambainis:degree-vs-qc, zhang:ambainis}] 
\label{thm:weighted}
Let $F: S \to \zo^m$ be a partial function.  Let $w, w'$ denote
a weight scheme as follows:
\begin{itemize}
\item Every pair $(x,y) \in S^2$ is assigned a non-negative weight
$w(x,y) = w(y,x)$ that satisfies $w(x,y) = 0$ whenever $F(x) = F(y)$.
\item Every triple $(x,y,i) \in S^2 \times [N]$ is assigned a
non-negative weight $w'(x,y,i)$ that satisfies $w'(x,y,i) = 0$
whenever $x_i = y_i$ or $F(x) = F(y)$, and $w'(x,y,i) w'(y,x,i) \ge
w^2(x,y)$ for all $x,y,i$ with $x_i \neq y_i$.
\end{itemize}
Then
\begin{equation*}
Q_2(F) = \Omega \Bigg(
  \min_{ \myatop{x,y,i}{\myatop{w(x,y) > 0} {x_i \ne y_i}}}
  \sqrt{\frac{wt(x) wt(y)}{v(x,i) v(y,i)}}
\ \Bigg),
\end{equation*}
where $wt(x) = \sum_y w(x,y)$ and $v(x,i) = \sum_y w'(x,y,i)$ for all
$x \in S$ and~$i \in [N]$.
\end{theorem}

At first glance, the weighted method may look rather complicated, both
in its formulation and use, though it is not.  We first assign weights
to pairs $(x,y)$ of inputs for which $F(x) \neq F(y)$, as in the
spectral method.  We require the weights to be symmetric so that they
represent the difficulty in distinguishing \emph{between $x$ and~$y$}.

We then afterwards assign weights $w'(x,y,i)$ that represent the
difficulty in distinguishing \emph{$x$ from $y$ by querying
index~$i$.}  The harder it is to distinguish $x$ from $y$ by
index~$i$, compared to distinguishing $y$ from $x$ by index~$i$, the
more weight we put on $(x,y,i)$ and the less on $(y,x,i)$, and
vice versa.  

To quantify this, define $t(x,y,i) = w'(x,y,i)/w'(y,x,i)$.  Then
$t(x,y,i)$ represents the relative amount of information we learn
about input pairs $(x,z)$ compared to the amount of information we
learn about input pairs $(u,y)$, by querying index~$i$.  If we, by
querying index~$i$, learn little about $x$ compared to $y$, we let
$t(x,y,i)$ be large, and otherwise small.  Consider we query an index
$i$ for which $x_i \neq y_i$.  Then we learn whether the oracle is $x$
or $y$.  However, at the same time, we also learn whether the oracle
is $x$ or $z$ for any other pair $(x,z)$ for which $x_i \neq z_i$ and
$F(x) \neq F(z)$; and similarly, we learn whether the oracle is $u$ or
$y$ for any other pair $(u,y)$ for which $u_i \neq y_i$ and $F(u) \neq
F(y)$.  The less information querying index $i$ provides about pairs
$(x,z)$ compared to pairs $(u,y)$, the larger we choose $t(x,y,i)$.
Having thus chosen $t(x,y,i)$, we set $w'(x,y,i) = w(x,y)
\sqrt{t(x,y,i)}$ and $w'(y,x,i) = w(x,y)/ \sqrt{t(x,y,i)}$.

We show next that the weighted method yields a lower bound of
$\Omega(\log N)$ for the ordered searching problem.  This proves that
the weighted method is strictly stronger than the unweighted method.
The weighted method yields strong lower bounds for read-once
formula~\cite{bs:q-read-once} and iterated
functions~\cite{ambainis:degree-vs-qc}.  Aaronson~\cite{aaronson:locsearch},
Santha and Szegedy~\cite{ss:local}, and Zhang~\cite{zhang:local} use
adversary arguments to prove lower bounds for local search, a
distributed version of Grover's problem.  {\v S}palek and Szegedy
prove in~\cite{ss:adversary} that the weighted method is equivalent to the
spectral method---any lower bound that can be achieved by one of the
two methods can also be shown by the other.  Their proof is
constructive and gives simple expressions for converting one into the
other.  The main weights $w(x,y)$ are the coefficients of the weight
function $W^t$ for the input pair $(x, y)$, that is, $w(x,y) =
\Gamma[x,y] \delta_x \delta_y$, and the secondary weights $w'(x,y,i)$
follow from Mathias' lemma~\cite{mathias:spectral-norm}
(Lemma~\ref{lm:mathias}).

\begin{example}
To apply the weighted method on ordered searching, we pick the same
weights $w(x,y) = \searchGamma[x,y]\, \delta_x \delta_y$ as in the
spectral method as there are no strong reasons for choosing otherwise.
Now, consider $t(x,y,i)$ with $F(x) \leq i < F(y)$ so that $x_i \neq
y_i$.  By querying index $i$, we also learn to distinguish between $x$
and $z$ for each of the $F(y)-i$ inputs $z$ with $i < F(z) \leq F(y)$,
and we learn to distinguish between $u$ and $y$ for each of the
$i-F(x)+1$ inputs $u$ with $F(x) \leq F(u) \leq i$.  We thus choose to
set
\begin{equation*}
t(x,y,i) = \frac{|F(x)-i| + 1}{|F(y)-i| + 1}.
\end{equation*}
Plugging these values into the weighted method yields a lower bound of
$\Omega(\log N)$ for ordered searching.
\end{example}

\section{Limitations of the spectral method}
\label{sec:limits}

The spectral method and the weighted adversary method
bound the amount of information that can be learned in any
one query.  They do not take into account that the amount of
information that can be learned in the \nth{j} query might differ from
the amount of information that can be learned in the \nth{k} query.

In~1999, Zalka~\cite{zalka:optimalgrover} successfully managed to
capture the amount of information that can be learned in each
individual query for a restricted version of Grover's
problem~\cite{grover:search}.  In this restricted version, we are
promised that the input oracle $x$ is either the zero-string (so
$|x|=0$) or exactly one entry in $x$ is one (so $|x|=1$), and the goal
is to determine which is the case.  By symmetry considerations, Zalka
demonstrates that Grover's algorithm saturates some improved
inequalities (which are similar to Eq.~\ref{eq:pair}) and hence is
optimal, even to within an additive constant.

Since current adversary methods do not capture the amount of
information the algorithm currently knows, we may simply assume that
the algorithm already knows every bit of the oracle and that it tries
to prove~so.  This motivates a study of the relationship between the
best bound achievable by the spectral method and the certificate
complexity.  A~\emph{certificate} for an input $x \in \zo^N$, is a
subset $C \subseteq [N]$ of input bits such that for any other input
$y$ in the domain of $F$ that may be obtained from $x$ by flipping
some of the indices not in~$C$, we have that $F(x) = F(y)$.  The
certificate complexity $\cert_x(F)$ of input $x$ is the size of a
smallest certificate for~$x$.  The \emph{certificate complexity}
$\cert(F)$ of a function $F$ is the maximum certificate complexity of
any of its inputs.  We also define the $z$-certificate complexity
$\cert_z(F)$ when taking the maximum only over inputs that map to~$z$.
The spectral theorem can then never yield a lower bound better than a
quantity that can be expressed in terms of certificate complexity.

\begin{lemma}
[\cite{lm:kolmogorov-lb, zhang:ambainis, ss:adversary}]
\label{lm:limit}
Let $F: S \rightarrow \zo$ be any partial boolean function.  The
spectral adversary lower bound $\adv(F)$ is at most $\min \big\{
\sqrt{ \cert_0(F) N}, \sqrt{ \cert_1(F) N} \big\}$.  If $F$ is total,
the method is limited by $\sqrt{ \cert_0(F) \cert_1(F)}$.
\end{lemma}

The certificate complexity of a function $F:\zo^N \rightarrow \zo^m$
is itself polynomially related to the block sensitivity of the
function.  An input $x \in \zo^N$ is \emph{sensitive} to a block $B
\subseteq [N]$ if $F(x) \neq F(x^B)$, where $x^B$ denotes the input
obtained by flipping the bits in $x$ with indices from~$B$.  The block
sensitivity $\bs_x(F)$ of input $x$ is the maximum number of disjoint
blocks $B_1, B_2, \ldots, B_k \subseteq [N]$ on which $x$ is
sensitive.  The \emph{block sensitivity} $\bs(F)$ of $F$ is the
maximum block sensitivity of any of its inputs.  We also define the
$z$-block sensitivity $\bs_z(F)$ when taking the maximum only over
inputs that map to~$z$.

For any boolean function $F: \zo^N \rightarrow \zo$, the certificate
complexity is upper-bounded by $\cert(F) \leq \bs_0 (F) \bs_1(F)$, and
thus so is the spectral adversary method.  Conversely, $\adv(F) \geq
\sqrt{\bs(F)}$ by a zero-one valued adversary matrix $\Gamma$: Let $x'
\in \zo^N$ be an input that achieves the block sensitivity of $F$, and
let $B_1, B_2, \ldots, B_k \subseteq [N]$ be disjoint blocks on which
$x'$ is sensitive, where $k = \bs(F)$.  Set $\Gamma(F)[x,x^B] = 1$ if
and only if $x = x'$ and $B$ is one of the $k$ blocks~$B_i$ and close
$\Gamma$ under transposition.  Then $\lambda(\Gamma) = \sqrt{k}$ and
$\max_i \lambda(\Gamma_i) = 1$, and thus
\begin{equation}
\sqrt{\bs(F)} \leq \adv(F) \leq \bs_0 (F) \bs_1(F).
\end{equation}

The spectral adversary method is not suitable for proving lower bounds
for problems related to property testing.  If function $F:S
\rightarrow \zo$ is a partial function with $S \subseteq \zo^N$ such
that every zero-input is of Hamming distance at least $\varepsilon n$
from every one-input, then the spectral theorem does not yield a lower
bound better than~$1/\varepsilon$.

Laplante and Magniez introduce in~\cite{lm:kolmogorov-lb} a
lower-bound method based on Kolmogorov complexity.  They show by
direct constructions that their method is at least as strong as each
of the two methods, the spectral and weighted adversary method.  {\v
S}palek and Szegedy then show in~\cite{ss:adversary} that the spectral
method is at least as strong as the Kolmogorov complexity method,
allowing us to conclude that the three methods are equivalent.  Having
such a variety of representations of the same method shows that the
adversary method is very versatile and captures fundamental properties
of functions.  Indeed, Laplante, Lee, and Szegedy~\cite{lls:formulas}
show that the square of the adversary bound is a lower bound on the
formula size.  The following lower-bound method is a combinatorial
version of the Kolmogorov complexity method.

\begin{theorem}
[Minimax method \cite{lm:kolmogorov-lb, ss:adversary}]
\label{thm:minimax}
Let $F: S \rightarrow \zo^m$ be a partial function and $\op{A}$ a
bounded-error quantum algorithm for~$F$.  Let $p: S \times [N]
\rightarrow \reals^{+}_{0}$ be a set of $|S|$ probability
distributions such that $p_x(i)$ denotes the average probability of
querying the \nth{i} input bit on input $x$, where the average is
taken over the whole computation of~$\op{A}$.  Then the query
complexity $Q_{\op{A}}$ of algorithm $\op{A}$ satisfies
\[
Q_{\op{A}} \geq M_p = \max_{ x,y: F(x) \ne F(y) }
  \frac{1}{\sum_{i: x_i \ne y_i} \sqrt{ p_x(i)\, p_y(i) }}.
\]
\end{theorem}

The previous methods satisfy the property that if we plug in some
matrix or relation, we get a valid lower bound.  The minimax method is
principally different.  A~lower bound computed by the minimax theorem
holds for one particular algorithm~$\op{A}$, and it may not hold for
some other and better algorithm.  However, we may obtain a universal
lower bound that holds for \emph{every} bounded error algorithm by
simply taking the minimum of the bound $M_p$ over all possible sets of
probability distributions~$p$.  The spectral bound and the minimax
bound are in a primal-dual relation: the best lower bound that can be
obtained by any adversary matrix $\Gamma$ equals the smallest bound
that can be obtained by a set of probability
distributions~$p$~\cite{ss:adversary}.  Primal methods are used for
obtaining concrete lower bounds and dual methods are used for proving
limitations of the method, as in Lemma~\ref{lm:limit}.

A useful property of the adversary method is that it composes.
Consider a function of the form $H = F \circ (G_1, \ldots, G_k)$,
where $F: \zo^k \rightarrow \zo$ and $G_i : \zo^{N_i} \rightarrow \zo$
for $i = 1, \ldots, k$ are partial boolean functions.  A~composition
theorem states the complexity of function $H$ in terms of the
complexities of $F$ and $G_1, \ldots, G_k$.  Barnum and
Saks~\cite{bs:q-read-once} use composition properties to prove a query
lower bound of $\Omega(\sqrt{N})$ for any read-once formula,
Ambainis~\cite{ambainis:degree-vs-qc} proves a composition lower bound
for iterated boolean functions, and Laplante, Lee, and
Szegedy~\cite{lls:formulas} prove a limitation on composition lower
bounds for functions $G_i$ for which the adversary bound is upper
bounded by a common bound~$b$.  To formulate a composition theorem for
arbitrary cases when the functions $G_i$ may have different adversary
bounds, we require a weighted version of the spectral method.

Let $F: \zo^N \rightarrow \zo$ be a partial boolean function and
$\alpha = (\alpha_1, \ldots, \alpha_N)$ a string of positive reals.
Let
\begin{equation*}
\adv_{\alpha}(F) = \max_\Gamma \min_i \left\{
  \alpha_i \frac{\lambda(\Gamma)} { \lambda(\Gamma_i) } \right\},
\end{equation*}
where $\Gamma$ ranges over all adversary matrices for~$F$.  If the weights
are all 1, then our new quantity $\adv_\alpha(F)$ coincides with the
spectral adversary bound and is thus a lower bound on the quantum query
complexity of $F$.  If the weights $\alpha$ are non-uniform, then
$\adv_\alpha(F)$ is a new abstract complexity measure that assigns cost
$\alpha_i$ to querying the \nth i input bit.  We can then
prove~\cite{hs:compose} that the quantity $\adv_{\alpha}$ composes in the
following sense.

\begin{theorem}
[Composition Theorem \cite{bs:q-read-once, ambainis:degree-vs-qc,
lls:formulas, hs:compose}] For any composite function $H = F \circ
(G_1, \ldots, G_k)$, where $F: \zo^k \rightarrow \zo$ and $G_i :
\zo^{N_i} \rightarrow \zo$ are partial boolean functions,
\begin{equation*}
\adv_\alpha(H) = \adv_\beta(F),
\end{equation*}
where $\beta_i = \adv_{\alpha^i}(G_i)$, and $\alpha = (\alpha^1,
\ldots, \alpha^k)$ is a $k$-tuple of strings $\alpha^i \in
{\reals^{+}}^{N_i}$.
\end{theorem}

A natural generalization of Grover's problem is the so-called $k$-fold
search problem in which we are promised that exactly $k$ entries of
the input oracle $x$ are one (so $|x|=k$), and the goal is to find all
of these $k$ indices.  We say an algorithm $\op{A}$ succeeds if it
outputs a subset $S \subseteq [N]$ of size $k$ and $S$ contains all
indices $i \in [N]$ for which $x_i=1$.  Thus, by definition, it fails
even if it outputs all but one of the $k$ indices.  The $k$-fold
search problem can be solved in $O(\sqrt{kn})$ queries, essentially by
sequentially running Grover's search algorithm $k$ times.  Klauck, {\v
S}palek, and de~Wolf~\cite{ksw:dpt} show that if the number of queries
is less than $\epsilon \sqrt{k n}$ for some constant~$\epsilon$, then
the success probability of $\op{A}$ is exponentially small in~$k$.
They thus prove a strong direct product theorem for the $k$-fold
search problem.  One of the main elements of the proof is the
polynomial method which we discuss in the next section.

In very recent work, Ambainis~\cite{ambainis:sdp} proposes an
extension of the adversary method and uses it to reprove the strong
direct product theorem of~\cite{ksw:dpt}.  Though the following very
brief description of the proof does not do full justice to the
method, we hope it conveys some of the intuition on
which~\cite{ambainis:sdp} is based.  The algorithm runs on a uniform
superposition of all inputs.  During the computation, the input
register gets entangled with the workspace of the algorithm due to the
queries to the oracle.  We trace out the workspace and examine the
eigenspaces of the density matrix of the input register.  Due to
symmetries, there are exactly $k+1$ eigenspaces, indexed by the number
of ones the algorithm ``knows'' at that stage of the algorithm.  In
the beginning, all amplitude is in the \nth 0 eigenspace.  One query
can only move little amplitude from the \nth i eigenspace to the
\nst{i+1} eigenspace.  If the algorithm has good success
probability, the quantum amplitude in high eigenspaces must be
significant, since the algorithm must ``know'' most of the $k$ indices, which
implies a lower bound on the query complexity.

\section{Polynomial lower bounds}
\label{sec:poly}

There are essentially two different methods known for proving lower
bounds on quantum computations.  The historically first method is the
adversary method we discuss above.  It was introduced in~1994 by
Bennett, Bernstein, Brassard, and Vazirani, and published in 1997 in
the SIAM Journal on Computing, in a special section that contains some
of the most outstanding papers on quantum computing.  The second
method was introduced shortly after, in 1998, by Beals, Buhrman,
Cleve, Mosca, and de~Wolf~\cite{bbcmw:polynomialsj}, and implicitly
used by Fortnow and Rogers in~\cite{fr:limits}.  Their approach is
algebraic and follows earlier very successful work on classical lower
bounds via polynomials (see for instance Beigel's 1993
survey~\cite{beigel:polmethod} and Regan's 1997
survey~\cite{regan:survey}).  We first establish that any partial
boolean function $F: S \rightarrow \zo$, where $S \subseteq \zo^N$,
can be represented by a real-valued polynomial $p: \reals^N
\rightarrow \reals$.

\begin{definition}
Let $F: S \to \zo$ be a partial boolean function, where $S \subseteq
\zo^N$.  An $N$-variable polynomial \emph{$p$ represents $F$} if $p(x)
= F(x)$ for all $x \in S$, and it \emph{approximates $F$} if $|p(x) -
F(x)| \leq \frac13$ for all $x \in S$.  The \emph{degree} of $F$,
denoted $\deg(F)$, is the minimal degree of a polynomial
representing~$F$.  The \emph{approximate degree} of $F$, denoted
$\adeg(F)$, is the minimal degree of a polynomial approximating~$F$.
\end{definition}

The crux in~\cite{bbcmw:polynomialsj} is in showing that any quantum
algorithm $\op{A}$ computing some function $F$ gives rise to some
polynomial~$p_{\op{A}}$ that represents or approximates~$F$.

\begin{theorem}[\cite{bbcmw:polynomialsj}]
\label{thm:poly}
Let $\op{A}$ be a quantum algorithm that computes a partial boolean
function~$F: S \rightarrow \zo$, where $S \subseteq \zo^N$, using at
most $T$ queries to the oracle~$\op{O}'_x$.  Then there exists an
$N$-variate real-valued multilinear polynomial $p_{\op{A}}: \reals^N
\rightarrow \reals$ of degree at most $2T$, which equals the
acceptance probability of~$\op{A}$.
\end{theorem}

\begin{proof}
In this theorem, we use the oracle $\op{O}'_x$ which is equivalent to
the oracle~$\op{O}_x$, since it allows for simple formulations.  We
first rewrite the action of $\op{O}'_x$ as
\begin{equation}\label{eq:onequerypoly}
\op{O}'_x \ket{i,b;z} 
 = (1 - x_i) \ket{i,b;z} + x_i \ket{i,b \oplus 1; z}
\end{equation}
where we define $x_i=0$ for $i=0$ so that we can simulate a non-query
by querying $x_i$ with $i=0$.  Suppose we apply $\op{O}'_x$ on some
superposition $\sum_{i,b,z} \alpha_{i,b,z} \ket{i,b;z}$ where each
amplitude $\alpha_{i,b,z}$ is an $N$-variate complex-valued polynomial
in $x$ of degree at most~$j$.  Then, by Eq.~\ref{eq:onequerypoly}, the
resulting state $\sum_{i,b,z} \beta_{i,b,z} \ket{i,b;z}$ is a
superposition where each amplitude $\beta_{i,b,z}$ is an $N$-variate
complex-valued polynomial in $x$ of degree at most $j+1$.  By~proof by
induction, after $T$ queries, each amplitude can be expressed as a
complex-valued polynomial in $x$ of degree at most~$T$.  The
probability that the final measurement yields the outcome~$1$,
corresponding to accepting the input, is obtained by summing some of
the absolute values of the amplitudes squared.  The square of any of
the absolute amplitudes can be expressed as a real-valued polynomial
$p_{\op{A}}$ in $x$ of degree at most~$2T$.  Theorem~\ref{thm:poly}
follows.
\end{proof}

The above theorem states that to any quantum algorithm $\op{A}$
computing a boolean function $F: S \rightarrow \zo$, where $S
\subseteq \zo^N$, we can associate an $N$-variate polynomial
$p_{\op{A}}: \reals^N \rightarrow \reals$ that expresses the
acceptance probability of the algorithm on any given input.  If
algorithm~$\op{A}$ is exact, i.e., if $\op{A}$ always stops and
outputs the correct answer, then $p_{\op{A}}(x)=F(x)$ for all $x \in
S$, and thus $p_{\op{A}}$ represents~$F$.  If $\op{A}$ has bounded
error, then $0 \leq p_{\op{A}}(x) \leq 1/3$ if $F(x)=0$ and $2/3 \leq
p_{\op{A}}(x) \leq 1$ if $F(x)=1$, and thus $p_{\op{A}}$
approximates~$F$.  The degree of $p_{\op{A}}$ is at most twice the
number of queries used by algorithm~$\op{A}$.  Consequently, the
degree of a function is a lower bound on the quantum query complexity,
up to a factor of two.

\begin{corollary}[Polynomial method \cite{bbcmw:polynomialsj}]
For any partial boolean function $F: S \rightarrow \zo$, where $S
\subseteq \zo^N$, we have $Q_E(F) \geq \deg(F)/ 2$ and $Q_2(F) \geq
\adeg(F) / 2$.
\end{corollary}

\section{Applying the polynomial method}
\label{sec:applyingpoly}

The challenge in applying the polynomial method lies in the
dimensionality of the input.  Typically, the method is applied by
first identifying a univariate or bivariate polynomial that captures
essential properties of the problem, and then proving a lower bound on
the degree of that polynomial.  The second part is typically
reasonably straightforward since polynomials have been studied for
centuries and much is known about their degrees.  The possibly
simplest nontrivial example is when $F$ is the threshold function
$\thr_t$ defined by $\thr_t(x) = 1$ if and only if $|x| \geq t$.  It
is easy to see that $\deg(\thr_t) = \Theta(N)$ for all nontrivial
threshold functions, and thus $Q_E(\thr_t) = \Omega(N)$.
Paturi~\cite{paturi:degree} shows that $\adeg(\thr_t) =
\Theta\big(\sqrt{(t+1)(N-t+1)}\big)$, and we thus readily get that
$Q_2(\thr_t) = \Omega\big(\sqrt{(t+1)(N-t+1)}\big)$, which is tight by
quantum counting~\cite{bhmt:countingj,bbcmw:polynomialsj}.  This degree
argument extends to any symmetric function~$F$ by writing $F$ as a sum
of threshold functions.  The same tight lower bounds for symmetric
functions can also be obtained by the unweighted adversary method (see
the paragraph after Theorem~\ref{thm:unweighted}).

For general non-symmetric functions, the polynomial method is,
however, significantly harder to apply.  For problems that are
``close'' to being symmetric, we can sometimes succeed in constructing
a univariate or bivariate polynomial that yields a non-trivial lower
bound.  The first and, in our view, most important such a result was
obtained by Aaronson in~\cite{aaronson:collision} in which he proves a lower
bound of $\Omega(N^{1/5})$ on any bounded-error quantum algorithm for
the collision problem.

The collision problem is a non-boolean promise problem.  The oracle is
an $N$-tuple of positive integers between 1 and~$M$, which we think of
as a function $X: [N] \rightarrow [M]$.  We model the oracle
$\op{O}''_X$ so that a query to the \nth{i} entry of the oracle
returns the integer~$X(i)$.  Specifically, $\op{O}''_X$ takes as input
$\ket{i,r;z}$ and outputs $\ket{i,r \oplus X(i);z}$ where $0 \leq r <
2^m$ for $m = \lceil \log_2(M+1) \rceil$, and $r \oplus X(i)$ denotes
bitwise addition modulo~2.  We are promised that either $X$ is a
one-to-one function, or $X$ is two-to-one, and the goal is to
determine which is the case.

The result of Aaronson was shortly after improved by
Shi~\cite{shi:collision} to~$\Omega(N^{1/4})$ for general functions
$X: [N] \rightarrow [M]$, and to $\Omega(N^{1/3})$ in the case the
range is larger than the domain by a constant factor, $M \geq
\frac{3}{2}N$.  The lower bounds of Aaronson and Shi appears as a
joint article~\cite{as:collision}.  Finally,
Kutin~\cite{kutin:collision} and Ambainis~\cite{ambainis:collision}
independently found remedies for the technical limitations in Shi's
proof, yielding an $\Omega(N^{1/3})$ lower bound for all functions,
which is tight by an algorithm that uses Grover search on subsets by
Brassard, H{\o}yer, and Tapp~\cite{bht:collision}.

The best lower bound for the collision problem that can be obtained
using the adversary method is only a constant, since any one-to-one
function is of large Hamming distance to any two-to-one function.
Koiran, Nesme, and Portier~\cite{knp:simon} use the polynomial method
to prove a lower bound of $\Omega(\log N)$ for Simon's
problem~\cite{simon:quantum}, which is tight~\cite{simon:quantum,bh:simon}.  
Simon's problem is a partial boolean function having properties related to
finite abelian groups.  Also for this problem, the best lower bound
that can be obtained using the adversary method is a constant.

In contrast, for any \emph{total} boolean function $F: \zo^N
\rightarrow \zo$, the adversary and polynomial method are both
polynomially related to block sensitivity,
\begin{gather}
\sqrt{\bs(F)/6} \leq \adeg(F) \leq \deg(F) \leq \bs^3(F)\\
\sqrt{\bs(F)} \leq \adv(F) \leq \bs^2(F).
\end{gather}
It follows from~\cite{bbcmw:polynomialsj} that $\deg(F) \leq
\bs^3(F)$, and from Nisan and Szegedy~\cite{nisan&szegedy:degree} that $6
\adeg(F)^2 \geq \bs(F)$.  Buhrman and
de~Wolf~\cite{buhrman&wolf:dectreesurvey} provide an excellent survey
of these and other complexity measures of boolean functions.

The polynomial lower bound is known to be inferior to the weighted
adversary method for some total boolean functions.
In~\cite{ambainis:degree-vs-qc}, Ambainis gives a boolean function $F:
\zo^4 \rightarrow \zo$ on four bits, which can be described as ``the
four input bits are sorted''~\cite{lls:formulas}, for which
$\deg(F)=2$ and for which there exists an adversary matrix~$\Gamma^F$
satisfying that $\lambda(\Gamma^F)/\max_i \lambda(\Gamma_i^F) = 2.5$.
We~compose the function with itself and obtain a boolean function
$F_{2} = F \circ (F,F,F,F): \zo^{16} \rightarrow \zo$ defined on 16
bits for which $\deg(F_{2})=4$, and for which
$\lambda(\Gamma^{F_{2}})/\max_i \lambda(\Gamma_i^{F_{2}}) = 2.5^2$, by
the composition theorem.  Iterating $n$ times, yields a function $F$
on $N=4^n$ bits of degree $\deg(F) = 2^n$, with spectral lower bound
$2.5^n = \deg(F)^{1.32\ldots}$, by the composition theorem.  The thus
constructed function $F$ is an example of an iterated function of low
degree and high quantum query complexity.  It is the currently biggest
known gap between the polynomial method and the adversary method for a
total function.  Another iterated total function for which the
adversary methods yield a lower bound better than the degree, is the
function described by ``all three input bits are
equal''~\cite{ambainis:degree-vs-qc}.

The polynomial method is very suitable when considering quantum
algorithms computing functions with error $\epsilon$ that is
sub-constant, whereas the adversary method is not formulated so as to
capture such a fine-grained analysis.  Buhrman, Cleve, de~Wolf, and
Zalka~\cite{bcwz:qerror} show that any quantum algorithm for Grover's
problem that succeeds in finding an index~$i$ for which $x_i=1$ with
probability at least $1-\epsilon$, provided one exists, requires
$\Omega(\sqrt{N \log(1/\epsilon)})$ queries to the oracle, which is tight.  
A~possibly more familiar example is that any polynomial approximating the 
parity function with any positive bias~$\epsilon>0$ (as opposed to bias
$\frac{1}{6}$ where $\frac{1}{6}=\frac{2}{3}-\frac{1}{2}$) has
degree~$N$, since any such polynomial gives rise to a univariate
polynomial of no larger degree with $N$ roots.  Hence, any quantum
algorithm computing the parity function with arbitrary small
bias~$\epsilon>0$ requires $N/2$ queries to the oracle, which is
tight.

A useful property of representing polynomials is that they compose.
If $p$ is a polynomial representing a function $F$, and polynomials
$q_1, q_2, \ldots, q_k$ represent functions $G_1, \ldots, G_k$, then
$p \circ (q_1, \ldots, q_k)$ represents $F \circ (G_1, \ldots, G_k)$,
when well-defined.  This composition property does not hold for
approximating polynomials: if each sub-polynomial $q_i$ takes the
value $0.8$, say, then we cannot say much about the value $p(0.8,
\ldots, 0.8)$ since the value of $p$ on non-integral inputs is not
restricted by the definition of being an approximating polynomial.  To
achieve composition properties, we require that the polynomials are
insensitive to small variations of the input bits.  Buhrman, Newman,
R{\"o}hrig, and de~Wolf give in~\cite{bnrw:robustq} a definition of
such polynomials, and refer to them as being robust.

\begin{definition}[Robust polynomials \cite{bnrw:robustq}] 
\label{def:robust}
An approximate $N$-variate polynomial $p$ is \emph{robust} on $S
\subseteq \zo^N$ if $|p(y) - p(x)| \leq \frac13$ for every $x \in S$
and $y \in \reals^M$ such that $|y_i - x_i| \leq \frac13$ for every
$i=1, \dots, M$.  The \emph{robust degree} of a boolean function $F: S
\rightarrow \zo$, denoted $\rdeg(F)$, is the minimal degree of a
robust polynomial approximating $F$.
\end{definition}

Robust polynomials compose by definition.  Buhrman
et~al.~\cite{bnrw:robustq} show that the robust degree of any total
function $F: \zo^N \rightarrow \zo$ is $O(N)$ by giving a classical
algorithm that uses a quantum subroutine for Grover's
problem~\cite{grover:search} which is tolerant to errors, due to
H{\o}yer, Mosca, and de~Wolf~\cite{hmw:berror-search}.  Buhrman
et~al.~\cite{bnrw:robustq} also show that $\rdeg(F) \in O(\adeg(F) \log
\adeg(F))$ by giving a construction for turning any approximating
polynomial into a robust polynomial at the cost of at most a
logarithmic factor in the degree of~$F$.  This implies that for any
composite function $H = F \circ (G, \ldots, G)$, we have $\adeg(H) \in
O(\adeg(F) \adeg(G) \log \adeg(F))$.  It is not known whether this is
tight.  Neither is it known if the approximate degree of $H$ can be
significantly smaller than the product of the approximate degrees of
$F$ and~$G$.  The only known lower bound on the approximate degree of
$H$ is the trivial bound $\Omega(\adeg(F) + \adeg(G))$.

An and-or tree of depth two is a composed function $F \circ (G,
\ldots, G)$ in which the outer function $F$ is the logical AND of
$\sqrt{N}$ bits, and the inner function $G$ is the logical OR of
$\sqrt{N}$ bits.  By the unweighted adversary method, computing and-or
trees of depth two requires $\Omega(\sqrt{N})$ queries.  H{\o}yer,
Mosca, and de~Wolf~\cite{hmw:berror-search} give a bounded-error
quantum algorithm that uses $O(\sqrt{N})$ queries, which thus is
tight.  The existence of that algorithm implies that there exists an
approximating polynomial for and-or tree of depth two of
degree~$O(\sqrt{N})$.  No other characterization of an approximating
polynomial for and-or trees of depth two of degree $O(\sqrt{N})$ is
currently known.  The best known lower bound on the approximate degree
of and-or trees of depth two is $\Omega(N^{1/3})$, up to logarithmic
factors in~$N$, by a folklore reduction from the element distinctness
problem on $\sqrt{N}$ integers~\cite{as:collision}.

\section{Concluding remarks}
\label{sec:concluding}

We have been focusing on two methods for proving lower bounds on
quantum query complexity: the adversary method and the polynomial
method.  Adversary lower bounds are in general easy to compute, but
are limited by the certificate complexity.  Known lower bounds are
constructed by identifying hard input pairs, finding weights
accordingly, and computing either the spectral norm of some matrices,
or applying the weighted method.  Polynomial lower bounds may yield
stronger bounds, but are hard to prove.  Known lower bounds by the
polynomial methods are constructed by identifying symmetries within
the problem, reducing the number of input variables to one or two, and
proving a lower bound on the degree of the reduced polynomial.

Barnum, Saks, and Szegedy give in~\cite{bss:semidef} a third lower
bound method that exactly characterizes the quantum query complexity,
but this strength turns out also to be its weakness: it is very hard
to apply and every known lower bound obtained by the method can also
be shown by one of the other two methods.  In a very recent work,
Ambainis~\cite{ambainis:sdp} extends the adversary method and uses it
to reprove a strong direct product theorem by Klauck, {\v S}palek, and
de~Wolf~\cite{ksw:dpt} obtained by techniques that include the
polynomial method.  Klauck et~al.~\cite{ksw:dpt} show that their
strong direct product theorem implies good quantum time-space
tradeoffs, including a quantum lower bound of $T^2 \cdot S =
\Omega(N^3)$ for sorting.  A~significant body of work have been
conducted on lower bounds on communication complexity.
We refer to de~Wolf's excellent
survey~\cite{wolf:qccsurvey} as a possible starting point.

There is a range of problems for which we do not currently know tight
quantum query bounds.  One important example is binary and-or trees of
logarithmic depth.  A binary and-or tree on $N=4^n$ variables is
obtained by iterating the function $F(x_1,x_2,x_3,x_4) = (x_1 \wedge
x_2) \vee (x_3 \wedge x_4)$ in total $n$ times.  The classical query
complexity for probabilistic algorithms is
$\Theta(N^{0.753})$~\cite{sw:and-or,snir:dec,santha:and-or}.  No
better bounded-error quantum algorithm is known.  The best known lower
bound on the quantum query complexity is $\Omega(\sqrt N)$ by
embedding the parity function on $\sqrt{N}$ bits and noting that the
parity function has linear query complexity, which can be shown by
either method.

Magniez, Santha, and Szegedy give in~\cite{mss:triangle} a quantum
algorithm for determining if a graph on $N$ vertices contains a
triangle which uses $O(N^{1.3})$ queries to the adjacency matrix.  The
best known lower bound is $\Omega(N)$ by the unweighted adversary
method, and has been conjectured not to be
tight~\cite{ambainis:degree-vs-qc}.  The problem of
triangle-identification is an example of a graph property, which is a
set of graphs closed under isomorphism.  Sun, Yao,
and~Zhang~\cite{syz:graphs} show that there exists a non-trivial graph
property of quantum query complexity $O(\sqrt{N})$, up to logarithmic
factors in~$N$.

Gasarch, in a survey on private information retrieval, published in
this Computational Complexity Column in the
Bulletin~\cite{gasarch:pir}, writes: ``A~field is interesting if it
answers a fundamental question, or connects to other fields that are
interesting, or uses techniques of interest.''  It~is our hope that
the reader will find that the surveyed area of quantum lower bounds
fulfills each of those three criteria.

\section*{Acknowledgments}

We thank Michal Kouck\'y and Kolja Vereshchagin for discussions on the
proof of the spectral adversary bound.


\newcommand{\etalchar}[1]{$^{#1}$}

\end{document}